# Low Complexity Beamforming Training Method for mmWave Communications

Felix Fellhauer*, Nabil Loghin†, Dana Ciochina†, Thomas Handte† and Stephan ten Brink*
*Institute of Telecommunications, Pfaffenwaldring 47, University of Stuttgart, 70569 Stuttgart, Germany
†Sony European Technology Center (EuTEC), Stuttgart, Germany

*Abstract*—This paper introduces a low complexity method for antenna sector selection in mmWave Hybrid MIMO communication systems like the IEEE 802.11ay amendment for Wireless LANs. The method is backwards compatible to the methods already defined for the released mmWave standard IEEE 802.11ad. We introduce an extension of the 802.11ad channel model to support common Hybrid MIMO configurations. The proposed method is evaluated and compared to the theoretical limit of transmission rates found by exhaustive search. In contrast to state-of-the-art solutions, the presented method requires sparse channel information only. Numerical results show a significant complexity reduction in terms of number of necessary trainings, while approaching maximum achievable rate.

## I. INTRODUCTION

The mmWave spectrum from $30\,\text{GHz}$ to $300\,\text{GHz}$ is essentially undeveloped for consumer telecommunications. As a consequence usage of these frequencies for future communication standards draws increasingly attention. The new spectrum comes with challenging propagation characteristics like strong attenuation by obstacles and huge path loss. To overcome path loss current standards such as [1] or [2] recommend the use of highly directive antennas such as Phased Array Antennas (PAAs). By means of efficient beamforming, data rates of up to multiple $^{\text{Gb}}/_{\text{s}}$ ($8085\,^{\text{Mb}}/_{\text{s}}$ for single-carrier in IEEE 802.11ad (11ad)) can be achieved in a Single Input Single Output (SISO) system.

Many modern communication systems use advanced technologies like Multiple Input Multiple Output (MIMO) in order to further increase spectral efficiency [3], [4]. Due to the fact that high cost and complexity are still considerable for mixed signal parts in mmWave spectrum, the introduction of a large number of Radio Frequency (RF) front ends for MIMO may not be cost-effective in mmWave communications. Further, low efficiency of mmWave mixed signal components poses an additional challenge for mobile devices in terms of power consumption and makes fully digital baseband precoding and combining for large numbers of spatial streams rather complex [5]. To find a trade-off between achievable spatial diversity, multiplexing gain and number of required RF front ends, the concept of Hybrid MIMO (H-MIMO) has been introduced [6]. H-MIMO allows a joint RF and baseband design with "soft" antenna selection using phase shifters connected to each antenna element. Thereby it allows to scale the cost and efficiency of a mmWave MIMO communication system, while still exploiting spatial diversity. Thus, the IEEE 802.11ay (11ay) amendment is going to make use of this hybrid architecture with up to 8 spatial streams, targeting transmission rates of multiples of $8085\,^{\text{Mb}}/_{\text{s}}$. The design of the phase shifter configurations can be formulated as a constrained optimization problem assuming perfect Channel State Information (CSI) at Receiver (RX) [7]. Achieving CSI in H-MIMO systems is a quite challenging task [8]. Furthermore, the large amount of feedback data imposes restrictions to real world implementations. Several approaches have been introduced to cope with this problem [9], [10]. We propose a heuristic approach to find phase shifter configurations without relying on complex H-MIMO channel estimation and solving the resulting computationally expensive optimization problem. The proposed method is divided into two stages, which ensures compatibility to the beam training frame structure of the well known 11ad standard. We use a standard indoor Channel Model (CM) to demonstrate that the proposed solution achieves rates close to the theoretical optimum under realistic conditions.

This paper is organized as follows. The H-MIMO system model for Transmitter (TX) and RX is introduced in Section II; specifically II-A and II-B explain the channel- and antenna model. In Section III we formulate the optimization problem for beam selection in H-MIMO systems. Section IV introduces a low complexity beam training algorithm together with an Exhaustive Search (ES) method that is used for benchmarking purposes. Simulation results and interpretation are presented in Section V, while Section VI gives some conclusions.

## II. SYSTEM MODEL

We employ the system model as visualized in Fig. 1a. For the ease of notation, we restrict ourselves to a H-MIMO system with identical configurations at TX and RX. This implies equal number of RF chains and equal configuration of all PAAs (referred to as symmetric). Further, we assume that two symbol streams $\mathbf{s}_1$ and $\mathbf{s}_2$ are transmitted simultaneously. These streams are fed into a digital precoder, described by precoding matrix $\mathbf{F}_{\text{BB}}$, and handed over to the analog part of the model which consists of $N_{\text{RF}} = 2$ RF chains followed by a bank of $N$ phase shifters for each PAA, described by the analog part of the precoding matrix $\mathbf{F}_{\text{RF}}$. The transmitted signals propagate through a wireless communication channel $\mathbf{H}_{\text{omni}}$, further described in II-A. At the RX side the same scheme follows the opposite order with $\mathbf{W}_{\text{BB}}$ and $\mathbf{W}_{\text{RF}}$, respectively. All elements of $\mathbf{F}_{\text{RF}}$ and $\mathbf{W}_{\text{RF}}$ are constrained by $|w_{\text{RF}}| = |f_{\text{RF}}| \equiv 1$, which results from the fact that phase shifters are passive components and, thus not able to control the signals amplitude. We further restrict the PAAs to employ a single spatial beam per PAA. Therefore, each PAA is modeled by the antenna model described in II-B.

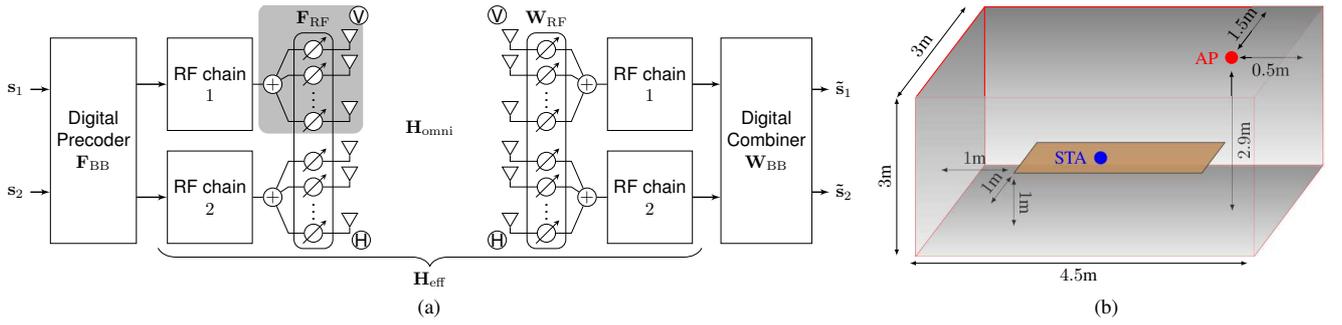

Fig. 1. (a): Hybrid MIMO communication system for transmitting two symbol streams $\mathbf{s}_1$ and $\mathbf{s}_2$ using two independent RF chains connected to two PAAs with distinct polarization (Vertical Ⓥ or Horizontal Ⓗ) and $N$ sub-antennas indicated by shaded area. (b): Conference Room scenario with single Station (STA) placed on a table and an Access Point (AP) mounted close to the ceiling [11].

### A. Channel Model

For simulations we use a version of the CM developed for the 11ad WLAN standard in 60 GHz domain, which follows a quasi deterministic approach [12]. We modify the CM to support H-MIMO by introducing polarization diversity. The signal propagation is modeled by rays connecting the TX and RX either directly or indirectly via reflections (deterministic part) and scattering caused by diffraction which forms a stochastically modeled cluster around each deterministic ray (stochastic part). The Channel Impulse Response (CIR) is generated in multiple steps, where cluster parameters are taken from ray-tracing generated Probability Density Functions (PDFs) for a given scenario (deterministic part) and blockage; intra cluster parameters, which are taken from a statistical model describing small scale effects like scattering. This approach leads to an analytical expression of a five-dimensional tapped delay line model for the channel (1) before analog beamforming, which we call omni-directional channel $\mathbf{H}_\text{omni}$.

$$\mathbf{H}_\text{omni} = \begin{pmatrix} h_{11} & h_{12} \\ h_{21} & h_{22} \end{pmatrix} \text{ with}$$
$$h_{nm} = h_{nm}(t, \varphi_\text{T}, \vartheta_\text{T}, \varphi_\text{R}, \vartheta_\text{R}) =$$
$$\sum_i a_{inm} C_i(t\text{-}T_i, \varphi_\text{T}\text{-}\phi_{\text{T},i}, \vartheta_\text{T}\text{-}\theta_{\text{T},i}, \varphi_\text{R}\text{-}\phi_{\text{R},i}, \vartheta_\text{R}\text{-}\theta_{\text{R},i})$$
$$\text{and } C_i(t, \varphi_\text{T}, \vartheta_\text{T}, \varphi_\text{R}, \vartheta_\text{R}) =$$
$$\sum_k \alpha_{i,k}\delta(t\text{-}\tau_{i,k})\delta(\varphi_\text{T}\text{-}\varphi_{\text{T},i,k})\delta(\vartheta_T\text{-}\vartheta_{\text{T},i,k})\ldots$$
$$\delta(\varphi_R\text{-}\varphi_{\text{R},i,k})\delta(\vartheta_R\text{-}\vartheta_{\text{R},i,k}) \quad (1)$$

Each element of the channel matrix $h_{nm}$ in (1) characterizes the CIR from TX-PAA $m$ to RX-PAA $n$. Each channel tap $k$ of cluster $i$ has the following properties: time of arrival $\tau$ (with respect to LOS-tap), complex valued amplitude $\alpha$, Angles of Departure (AoD) in azimuth $\varphi_\text{T}$ and elevation $\vartheta_\text{T}$ as well as respective Angles of Arrival (AoA) $\varphi_\text{R}$ and $\vartheta_\text{R}$. As it can be seen from (1) the channel is modeled on cluster-level (deterministic part) with $T_i$, $\phi_{\text{T},i}$, $\theta_{\text{T},i}$, $\phi_{\text{R},i}$ and $\theta_{\text{R},i}$ describing the time and angular properties of the $i$-th cluster. And on ray-level (stochastic part) with $\tau_{i,k}$, $\varphi_{\text{T},i,k}$, $\vartheta_{\text{T},i,k}$, $\varphi_{\text{R},k}$ and $\vartheta_{\text{R},i,k}$ the relative time and angular properties of ray $k$ within cluster $i$. As the polarization characteristics are almost equal within a cluster of rays, polarization is modeled on cluster level by applying a polarization specific gain $a_{inm}$ depending on the clusters polarization properties and respective antenna polarization. So both of the PAAs observe the same $\mathbf{H}_\text{omni}$ except for varying ray amplitudes due to polarization. The influence of polarization is modeled by Jones vectors depicting the antenna elements' orientation of each PAA, either linearly polarized in $\varphi$ direction ($\mathbf{e}_\text{v} = (0,\ 1)^\text{T}$) or in $\vartheta$ direction ($\mathbf{e}_\text{h} = (1,\ 0)^\text{T}$). Then, the attenuation coefficient matrix due to polarization $\mathbf{A}_i$ can be calculated independently for each cluster $i$. PDFs for each possible cluster type within the scenario have been generated by ray-tracing and are used to determine actual entries of the channel polarization matrix $\mathbf{H}_{i,\text{pol}}$. The components of each polarization matrix $\mathbf{H}_{i,\text{pol}}$ can be interpreted as gain coefficients between the horizontal and vertical E-field components at the TX and RX antenna.

$$\mathbf{A}_i = \begin{pmatrix} \mathbf{e}_\text{v}^\text{H}\mathbf{H}_{i,\text{pol}}\mathbf{e}_\text{v} & \mathbf{e}_\text{v}^\text{H}\mathbf{H}_{i,\text{pol}}\mathbf{e}_\text{h} \\ \mathbf{e}_\text{h}^\text{H}\mathbf{H}_{i,\text{pol}}\mathbf{e}_\text{v} & \mathbf{e}_\text{h}^\text{H}\mathbf{H}_{i,\text{pol}}\mathbf{e}_\text{h} \end{pmatrix} \quad (2)$$

We use the Conference Room (CR) scenario, which is visualized in Fig. 1b for uniformly distributed locations of the STA in the shaded area which emulates a table. In our setup we assume a H-MIMO configuration, TX and RX consisting of two cross-polarized PAAs without spatial separation (referring to SU-MIMO configuration #2 in 11ay [11]).

Thus, the CM describes the method for finding an omni directional CIR $\mathbf{H}_\text{omni}$. To further model the impact of the analog preceding and combining on the channel, we introduce an antenna model that utilizes spatial filtering in the following Section. This procedure derives the effective channel impulse response $\mathbf{H}_\text{eff}$ in time domain, which is finally observed at the digital part. Discretization with sampling frequency $f_s = 2.56\,\text{GHz}$ is us giving $h_{k,nm,\text{eff}}$ from $h_{nm,\text{eff}}(t)$.

Despite the fact that the Orthogonal Frequency Division Multiplex (OFDM) part of the 11ad standard has been deprecated, we use the original specification with with $N_\text{sub} = 512$ subcarriers for the purpose of benchmarking. Therefore we define the Discrete Fourier Transform of the baseband channel impulse response as $\mathbf{H}_{n,\text{eff}} = \mathcal{F}_N\{\mathbf{H}_{k,\text{eff}}\}$ using $N_\text{sub}$ subcarriers which will be referred to as "effective transfer function".

## B. Antenna Model

Usually phased array antennas applied in mmWave systems are either uniform linear arrays (ULAs) with antenna elements arranged in a single dimension with uniform spacing, or two-dimensional antenna arrays with uniform spacing in $x$ and $y$ dimensions, i.e. so called uniform rectangular arrays (URAs). The process of calculating the antenna gain pattern for each desired direction of the main lobe is computationally very expensive and thus simplified as described within the following paragraphs: For the purpose of simulation, we use a simplistic approach that applies a Gaussian antenna gain pattern in linear scale like in [12]. For all results in this paper the half power beam width is set to $\vartheta_{-3\mathrm{dB}} = 60°$. Fig. 2 shows an example of respective antenna gain patterns for an PAA with four elements and a Gaussian antenna gain pattern with a half power beam width of $30°$.

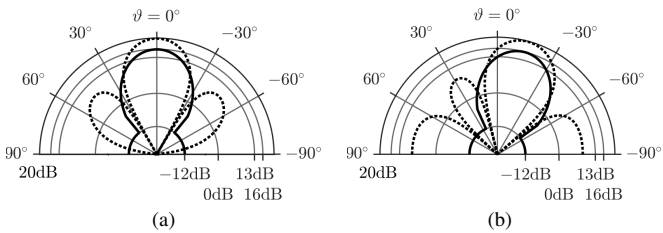

Fig. 2. Comparison of Gaussian antenna gain pattern (solid line) and four element PAA (dashed line). With main lobe direction $\theta_0 = 0°$ (a) and $\theta_0 = 15°$ (b).

To derive the respective codebook $\mathcal{C}$, which is a set of polar directions $c_i = (\vartheta_i, \varphi_i)$, a rotation symmetric spherical sector of width of $\vartheta_{\mathrm{sec}} = 90°$ is covered with hexagons evenly. All directions that correspond to center points belong to the codebook [12]. This results in $\mathcal{C} = \{c_1, ..., c_\ell\}$ of length $\ell = 19$. To simplify notation we further define the vectors $\mathbf{i}^N$ and $\mathbf{j}^N$ that contain the selected codebook indices at TX and RX side. Thus, for a $2 \times 2$ H-MIMO setup, (3) shows this relation. A codebook index fully specifies the spatial direction of the antennas' main lobe.

$$i_1 \to c_i = (\phi_{0,\mathrm{T}1}, \theta_{0,\mathrm{T}1}), i_2 \to c_i = (\phi_{0,\mathrm{T}2}, \theta_{0,\mathrm{T}2})$$
$$j_1 \to c_j = (\phi_{0,\mathrm{R}1}, \theta_{0,\mathrm{R}1}), j_2 \to c_j = (\phi_{0,\mathrm{R}2}, \theta_{0,\mathrm{R}2}) \quad (3)$$

To derive the effective channel impulse response based on the omni directional CIR (see (1)), the gain pattern of the selected codebook entry for each PAA is applied. Thus each tap amplitude gets amplified or attenuated accordingly to its AoA, AoD and selected codebook entry. (4) states this relation by applying spatial filtering on (1) [12].

$$h_{nm,\mathrm{eff}} = \int_0^{2\pi}\int_0^{\pi}\int_0^{2\pi}\int_0^{\pi} g_{\mathrm{R}}(\varphi_{\mathrm{R}}\text{-}\phi_{0,\mathrm{R}n}, \vartheta_{\mathrm{R}}\text{-}\theta_{0,\mathrm{R}n}) \cdot$$
$$h_{nm}(t, \varphi_{\mathrm{T}}, \vartheta_{\mathrm{T}}, \varphi_{\mathrm{R}}, \vartheta_{\mathrm{R}}) \cdot g_{\mathrm{T}}(\varphi_{\mathrm{T}}\text{-}\phi_{0,\mathrm{T}m}, \vartheta_{\mathrm{T}}\text{-}\theta_{0,\mathrm{T}m}) \cdot$$
$$\sin(\vartheta_{\mathrm{R}}) \cdot d\vartheta_{\mathrm{R}}d\varphi_{\mathrm{R}} \sin(\vartheta_{\mathrm{T}}) \cdot d\vartheta_{\mathrm{T}}d\varphi_{\mathrm{T}} \quad (4)$$

In addition to the directional antenna introduced before, the 11ad standard defines a second, quasi-omni antenna type [13]. This second antenna type allows to apply a directive antenna only on TX side of the communication system, while the RX listens omni-directionally. In order to model this, we introduce the constant gain pattern in all directions:

$$g_{\mathrm{R}}(\varphi_{\mathrm{R}}, \vartheta_{\mathrm{R}}) = g_0 \equiv 1 \quad (5)$$

## III. PROBLEM STATEMENT

The goal of the beam training procedure can be formulated as optimization problem. Hence to find the codebook entries for all PAAs specified by index vectors $\mathbf{i}$ and $\mathbf{j}$ that allow the highest rate assuming Singular Value Decomposition (SVD) per OFDM subcarrier to be performed in baseband signal processing. This relation is shown in (6) by using (3) and (4), for a given average SNR per receive PAA denoted as $\rho$.

$$\arg\max_{\forall \mathbf{i},\mathbf{j}}\left\{\sum_{n=1}^{N} \log_2 \det\left(\mathbf{I}_N + \mathbf{H}_{n,\mathrm{eff}}(\mathbf{i},\mathbf{j})\mathbf{H}_{n,\mathrm{eff}}^{\mathrm{H}}(\mathbf{i},\mathbf{j})\rho N_{\mathrm{RF}}\right)\frac{1}{N_{\mathrm{sub}}}\right\} \quad (6)$$

Due to the optimization over codebook entries, finding the global optimum of highest rate has exponential complexity. This is prohibitive in practice. Therefore our focus is to develop a reduced search method. Conventional MIMO methods cannot be applied straight forward, as it is tedious to estimate the full MIMO channel between all sub-antennas in practice [8]. Common solutions like SVD that provide a one-shot solution, are only applicable to the digital part of the system.

For the sake of simplicity, beam training might be performed as defined in the 11ad SISO-standard by treating both direct links $h_{11}$ and $h_{22}$ independently, while ignoring interference links $h_{12}$ and $h_{21}$. For the sake of completeness, the signaling to perform beam training in 11ad is described in IV-A. We further show in Section V that this approach leads to significant loss in performance.

## IV. EVALUATED METHODS

In our approach we make use of the 11ad beam training procedure, which we briefly remind in IV-A. Since this was only developed for SISO schemes, we show our proposed extension to MIMO in section IV-C. The performance of our proposed approach is then evaluated in Section V. As an upper bound, the ES method, as described in IV-B is implemented. Although it is not practical to search all combinations of codebook entries exhaustively for the global rate maximum it still provides an upper bound to benchmark less complex methods. As a practical solution, the Pairwise Search (PS) method has been proposed in order to reduce training iterations but is not further discussed here [9].

### A. SISO Beam Training in IEEE 802.11ad

The beam training procedure in 11ad is comprised of two stages, namely Sector Level Sweep (SLS) and Beam Refinement Phase (BRP) [1]. Both phases are performed pairwise between two stations (STA-A, STA-B), or a station and an AP, where the device that transmits first is called the Initiator, the second the Responder. During the first phase RX side is listening with quasi-omni-directional antenna gain patterns, here only information regarding the received signal quality (SNR) is exchanged. SLS can be followed by BRP, which intends to overhaul the deviations caused by inhomogeneity of quasi-omni receive patterns by testing directive beams at both sides, followed by signal quality and sector ID feedback. It can be shown, that when directly applied to MIMO channels, there is a huge loss in resulting spectral efficiency compared to the theoretical limit.

*B. MIMO Exhaustive Search*

As a straightforward approach to find the best channel formed by a set of discrete codebook entries at TX and RX one can systematically evaluate all possible combinations. For a symmetric H-MIMO system with same codebook length for all analog precoders and number of RF chains $N_{\text{RF}}$ at TX and RX this approach results in the total number of training iterations $N_{\text{tot,ES}} = \ell^{2N}$. This is not applicable to real world implementations due to the high number of channel estimations that has to be done. Nevertheless the algorithm guarantees to find the globally best codebook entries for maximizing the H-MIMO channel rate and, therefore, can be used to find the theoretical upper bound for benchmarking other algorithms.

*C. K-Best Algorithm*

With the goal of further reducing the number of required channel estimations we introduced a heuristic method [14], that is designed to cope with the beam training and feedback methods defined in the 11ad standard. Therefore, the procedure is divided into two stages:

*1) SISO-Phase:* The first phase intends to find spatial directions or beams providing strong direct paths or reflections as good candidates to form a high-rate MIMO channel, assuming that pairing of links applying equal polarization is predetermined. This is done by creating a beam score vector $\mathbf{q}_{n,\text{I/R}}$ for each of the $n = 1 \ldots N_{\text{RF}}$ PAAs and for Initiator and Responder, consisting of values indicating the received signal strength of the resulting beam-to-omni channel. The beam score vectors consist of received signal power values of the different effective channels. This metric is chosen because it can be measured with low effort and is available as Received Signal Strength Indicator (RSSI) in most RX architectures anyways. This relation is stated by using expression (5) in (4) which leads to (7), where $\mathbf{q}_{n,\text{I}}(i_n)$ denotes the RSSI at the $n$-th Responder PAA of the beam corresponding to codebook entry $c_{i_n}$ sent by the $n$-th Initiator PAA.

$$\mathbf{q}_{n,\text{I/R}} = \sum_t \left| \int_0^{2\pi} \int_0^{\pi} g_{\text{T}}(i_n/j_n) \cdot \\ h_{nn}(t, \varphi_{\text{T}}, \vartheta_{\text{T}}, \varphi_{\text{R}}, \vartheta_{\text{R}}) \cdot \sin(\vartheta_{\text{T}}) \cdot d\vartheta_{\text{T}} d\varphi_{\text{T}} \right|^2 \quad (7)$$

*2) MIMO-Phase:* In the second phase, the goal is to find the effective channel that finally maximizes the MIMO rate as stated in (6). Therefore, we propose a reduced search approach in which only a subset of beam combinations from the first stage are used. The subset of beam combinations is derived from the beam score vectors $\mathbf{q}_{n,\text{I/R}}$ by searching for the $K$ best combinations $b_k$ of products of beam score entries (joint beam score). Thus, each entry of the resulting sorted vector $\mathbf{b}$ of length $K$ is an indication of quality of the corresponding effective MIMO channel. This is done by finding matrices of indices $\hat{\mathbf{I}}^{(N \times K)}, \hat{\mathbf{J}}^{(N \times K)}$ that fulfill following condition:

$$b_k = \prod_{n=1}^{N} q_{n,\text{I}}(\hat{j}_{nk}) q_{n,\text{R}}(\hat{i}_{nk}) \text{ that } b_1 \geq b_2 \geq \cdots \geq b_K$$

with index sets $\hat{\mathbf{I}}$ and $\hat{\mathbf{J}}$ from $i,j \in \{1, 2, \ldots, l\}$ (8)

While the best candidate can easily be found by combining all best beam score values, all subsequent candidates in (8) require much more calculations. This can be achieved in a quite efficient manner but is not addressed in this paper. In contrast to the *SISO-Phase in 1)* we now assume full CSI of the effective MIMO channel $\mathbf{H}_{\text{eff}}$ which is necessary in order to estimate the respective MIMO rate.

After evaluating the MIMO rate for each of the $K$ combinations, the best combination is chosen to form the final effective channel. This simplifies the expression of the overall rate goal in (6) into (9).

$$\arg\max_{\forall k} \left\{ \sum_{n=1}^{N} \log_2 \det \left( \mathbf{I}_N + \\ \mathbf{H}_{n,\text{eff}}(\hat{\mathbf{i}}_k, \hat{\mathbf{j}}_k) \mathbf{H}_{n,\text{eff}}^{\text{H}}(\hat{\mathbf{i}}_k, \hat{\mathbf{j}}_k) \rho N_{\text{RF}} \right) \frac{1}{N_{\text{sub}}} \right\} \quad (9)$$

This results in a total number of iterations that is linearly dependent on the total number of RF chains $N_{\text{RF}}$ and codebook length $\ell$ in the first stage, plus the $K$ candidates which can be tested in a refinement phase similar to the one defined in the 11ad standard.

$$N_{\text{tot,KB}} = 2N_{\text{RF}} \cdot \ell + K \quad (10)$$

With $K$ left as an implementation specific parameter, $N_{\text{tot,KB}} \ll N_{\text{tot,ES}}$ holds for realistic values of $K$ and $\ell$. The impact of this parameter is characterized in more detail in the following Section.

V. NUMERICAL RESULTS

To characterize the proposed $K$-Best method, we simulate a set of 50 channel realizations in the Conference Room CM scenario described in II-A.

For each channel realization the STA location is assumed to be randomly selected from a uniform distribution bounded by the area of the table. The STA's angular orientation in the horizontal plane is also uniformly distributed. We assume a symmetric H-MIMO configuration with $N_{\text{RX}}^{\text{RF}} = N_{\text{TX}}^{\text{RF}} = 2$. Location of both PAAs at STA- and AP-side are equal thus no spatial diversity is utilized. Further, the two streams are assumed to be radiated by cross-polarized antennas which introduces some suppression of cross-links. This corresponds to the SU-MIMO Configuration #2 in [11]. The beam training procedure is most challenging in non-line-of-sight (NLOS) situations and thus, we assume NLOS for all channel realizations by removing the LOS-tap from all CIRs. The employed codebook of length $\ell = 19$ is derived by the antenna model introduced in II-B with an half power beamwidth of $\theta_{-3dB} = 60°$. Simulation parameters are summarized in Table I.

TABLE I

| Parameter | Value |
|---|---|
| Scenario | CR STA-AP |
| PAA Separation | $d = 0$ |
| Transmission Type | NLOS |
| Half Power Beam Width | $\theta_{-3dB} = 60°$ |
| SNR per RX PAA | $\rho = 20$ dB |
| Codebook Length | $\ell = 19$ |
| Channel Realizations | 50 |
| RF chains | $N_{RF} = 2$ |
| Investigated Combinations | $K = 100$ |

Fig. 3 shows the respective results. We compare the average channel rate that can be reached by the ES method to the rate achieved with $K$-Best with different values for $K$. All rate values are evaluated for an SNR value of $\rho = 20$ dB per receive antenna array and normalized to the rate, that can be achieved by the ES method. We also show results for the less complex PS method [9], reaching same performance as ES.

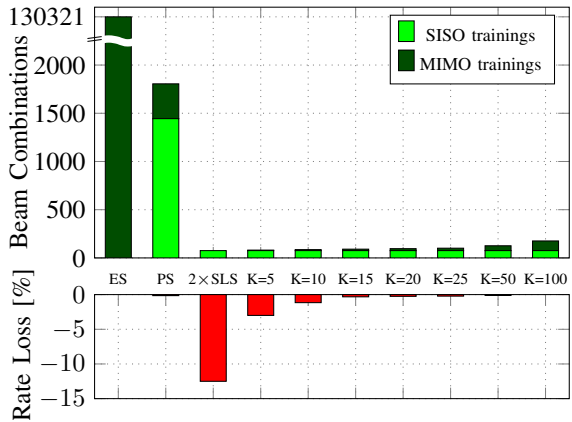

Fig. 3. Comparison of numerical results in the conference room indoor scenario for evaluated beam training methods. In the upper part the complexity of the methods is shown in terms of necessary training iterations. Lower part shows resulting relative rate loss compared to exhaustive search method.

It can be seen that the $K$-Best algorithm significantly reduces complexity in terms of total number of necessary training iterations. Especially the number of time consuming MIMO trainings can be reduced (full MIMO channel estimation required). Results show reduction of MIMO training iterations in the range of tree orders of magnitude compared to the ES method, while rates converge within the first 10 to 15 iterations of the $K$-Best algorithm. Further results reveal that SISO training (2×SLS) leads to noticeable losses in achievable rates ($\approx$ -12%). This might be for two reasons: First, there is no MIMO metric considered. Second, the beam-to-beam phase allows to compensate for misalignment of discrete beam directions resulting from the beam-to-omni simplification in first phase.

## VI. CONCLUSIONS

In this paper we introduced a low complexity beam training method for H-MIMO configurations that in contrast to other methods [8] only relies on signal strength measurements and comparably low number of measurements of the effective baseband channel $\mathbf{H}_{eff}$. The method is particularly well suited for systems that offer limited computation performance and limited CSI-feedback. Further, it is designed to cope with the beam training framing in the mmWave WLAN standard 11ad, and is thus a suitable approach for the upcoming standard 11ay featuring H-MIMO architectures.

The proposed method is evaluated using a version of the 11ad channel model for a 60 GHz indoor channel, modified to model polarization diversity between direct H-MIMO paths.

Simulation results show that performing beam training on SISO channels only, results in rate losses $> 10\%$. Results also show that by systematically testing through only the $K \gtrsim 10$ best SISO codebook combinations in a second stage, rate-loss can be reduced below $1\%$.